# Thermodynamic and magnetic properties of the hexagonal type Ising nanowire

Yusuf Kocakaplan[a] and Ersin Kantar[b,*]

[a]Graduate School of Natural and Applied Sciences, Erciyes University, 38039 Kayseri, Turkey
[b]Department of Physics, Erciyes University, 38039 Kayseri, Turkey

**Abstract**

The thermodynamic and magnetic properties of the mixed spin (1/2-1) hexagonal Ising nanowire (HIN) system with core-shell structure have been presented by means of the effective-field theory (EFT) with correlations. The effects of the physical parameters of the system on thermodynmaic and magnetic properties (magnetizations, susceptibilities, internal energies, and free energies and hysteresis curves) are investigated for both ferromagnetic and antiferromagnetic case, in detail. One can find that when the temperature increases the hysteresis loop areas decrease and the hysteresis loops disappear at above critical temperature. Moreover, different hysteresis loop behaviors have been observed such as single, double and triple hysteresis loops in the system. In order to confirm the accuracy of the phase transition points, we also investigate the free energy of the system.

## 1. Introduction

Recently, there is a continuously increasing interest in nanomaterials with core-shell structure owing to the fact that these materials have prospective applications in diverse field such as magnetic resonance imaging, cell and DNA separation, drug delivery [1], ultra-high density magnetic recording media [2-4], sensor [5], environmental remediation [6], nonlinear optics [7], bio-molecular motor [8] and permanent magnets [9]. Furthermore, nanomaterials have been studied within various model both experimentally and theoretically [10-22].

It is worth noting that thermodynamic and magnetic properties of nano materials have been investigated by using the Ising model and its variants. On the other hand, mixed spin Ising systems provide good models to investigate ferrimagnetism. One of the earliest, simplest and as well as most extensively studied mixed spin Ising model is the spin-1/2 and spin-1 mixed system. This system has been studied by the mean-field approximation (MFA) [23–25], Monte Carlo simulation (MCs) [26, 27] and EFT [28–32]. Moreover, the mixed spin (1/2-1) Ising model has been investigated on the nanostructures within the framework of the EFT with correlations [33-35].

---

[*]Corresponding author.
 Tel: + 90 352 207 66 66 # 33136; Fax: + 90 352 4374931.
 E-mail address: ersinkantar@erciyes.edu.tr  (E. Kantar)

We should also mention that magnetic hysteresis behaviors have been one of the most important and interesting symbol of magnetism in materials. Especially, the hysteresis behaviors have been studying for nanomaterials from both the experimental and theoretical points of view. Experimentally, the hysteresis behaviors have been studied for the ferromagnetism in nanowires [36, 37], in La$_{2/3}$Sr$_{1/3}$MnO$_3$ nanoparticle assembled nanotubes [38] and for carbon nanotubes [39-42]. Theoretically, the hysteresis behaviors have been investigated the ferromagnetic single-walled nanotubes [43], nanomagnets [44] and noninteracting nanoparticles [45].

Despite these studies, as far as we know, the thermodynamic and magnetic properties and hysteresis behaviors of mixed spin (1/2-1) HIN system have not been investigated. Therefore, in this paper, the effects of the physical parameters of the system on thermodynamic and magnetic properties and the hysteresis behaviors are studied for both ferromagnetic and antiferromagnetic case, in detail.

The paper is organized as follows. In Section 2, the model and formalism of the EFT with correlations is presented briefly. The detailed numerical results and discussions are given in Section 3. Finally, Section 4 is devoted to a summary and a brief conclusion.

**2. Model and formulation**

The Hamiltonian of the hexagonal Ising nanowire (HIN) includes nearest neighbors interactions and the crystal field is given as:

$$H = -J_S \sum_{\langle ij \rangle} S_i S_j - J_C \sum_{\langle mn \rangle} \sigma_m \sigma_n - J_1 \sum_{\langle im \rangle} S_i \sigma_m - D \sum_i S_i^2 - h \left( \sum_i S_i + \sum_m \sigma_m \right) \quad (1)$$

where $\sigma = \pm 1/2$ and $S = \pm 1, 0$. The $J_S$, $J_C$ and $J_1$ are the exchange interaction parameters between the two nearest-neighbor magnetic particles at the shell surface, core and between shell surface and core, respectively (see Fig. 1). D is Hamiltonian parameter and stand for the single-ion anisotropy (i.e. crystal field). The surface exchange interaction $J_S = J_C (1 + \Delta_S)$ and interfacial coupling $r = J_1 / J_C$ are often defined to clarify the effects of the surface and interfacial exchange interactions on the physical properties in the nanosystem, respectively.

For the mixed spin (1/2-1) HIN system, within the framework of the EFT with correlations, one can easily find the magnetizations and the quadruple moment as coupled equations as follows:

$$m_S = \left[1 + m_S \sinh(J_S \nabla) + m_S^2 (\cosh(J_S \nabla) - 1)\right]^4 \left[\cosh(J_I \nabla / 2) + 2m_C \sinh(J_I \nabla / 2)\right] F_1(x)\big|_{x=0}, \quad (2a)$$

$$m_C = \left[\cosh(J_C \nabla / 2) + 2m_C \sinh(J_C \nabla / 2)\right]^2 \left[1 + m_S \sinh(J_I \nabla) + m_S^2 (\cosh(J_I \nabla) - 1)\right]^6 G(x)\big|_{x=0}, \quad (2b)$$

$$q_S = \left[1 + m_S \sinh(J_S \nabla) + m_S^2 (\cosh(J_S \nabla) - 1)\right]^4 \left[\cosh(J_I \nabla / 2) + 2m_C \sinh(J_I \nabla / 2)\right] F_2(x)\big|_{x=0}, \quad (2c)$$

where $\nabla = \partial / \partial x$ is the differential operator. The functions $F_1(x)$, $F_2(x)$ and $G(x)$ are defined as

$$F_1(x) = \frac{2\sinh[\beta(x+h)]}{\exp(-\beta D) + 2\cosh[\beta(x+h)]} \quad (3a)$$

$$F_2(x) = \frac{2\cosh[\beta(x+h)]}{\exp(-\beta D) + 2\cosh[\beta(x+h)]} \quad (3b)$$

$$G(x) = \frac{1}{2} \tanh\left[\frac{1}{2}\beta(x+h)\right]. \quad (3c)$$

Here, $\beta = 1/k_B T$, T is the absolute temperature and $k_B$ is the Boltzmann constant and it is selected as $k_B = 1.0$ during the paper. By using the definitions of the order parameters in Eqs. (2a)-(2c), the total $(M_T)$ magnetizations of per site can be defined as $M_T = 1/7(6m_S + m_C)$.

In order to obtained susceptibilities of the system, we differentiated magnetizations respect to h as following equation:

$$\chi_\alpha = \lim_{\to 0} \left(\frac{\partial m_\alpha}{\partial h}\right) \quad (4)$$

where, $\alpha = C$ and S. By using of Eqs. (2) and (4), we can easily obtain the $\chi_C$ and $\chi_S$ suscebtibilites as follow:

$$\chi_C = a_1 \chi_C + a_2 \chi_S + a_3 \frac{\partial F_1(x)}{\partial h}, \tag{5a}$$

$$\chi_S = b_1 \chi_S + b_2 \chi_C + b_3 \frac{\partial F_2(x)}{\partial h}. \tag{5b}$$

Here, $a_i$ and $b_i$ (i=1, 2 and 3) coefficients have complicated and long expressions, hence they will not give. The total susceptibilities of per site can be obtain via $\chi_T = 1/7(\chi_C + 6\chi_S)$.

The internal energy of per site of the system can be calculated as

$$\frac{U}{N} = -\frac{1}{2}(\langle E_{m_c} \rangle + \langle E_{m_s} \rangle) - h(m_C + m_S) - D(q_S), \tag{6}$$

where,

$$\langle U_C \rangle = \frac{\partial}{\partial \nabla}[\cosh(J_C \nabla / 2) + 2m_C \sinh(J_C \nabla / 2)]^2 \\ \times [1 + m_S \sinh(J_1 \nabla) + m_S^2(\cosh(J_1 \nabla) - 1)]^6 G(x+h)\Big|_{x=0}, \tag{7a}$$

$$\langle U_S \rangle = \frac{\partial}{\partial \nabla}[1 + m_S \sinh(J_S \nabla) + m_S^2(\cosh(J_S \nabla) - 1)]^4 \\ \times [\cosh(J_1 \nabla / 2) + 2m_C \sinh(J_1 \nabla / 2)] F_1(x+h)\Big|_{x=0}. \tag{7b}$$

The specific heat of the system can be obtained from the relation

$$C_h = \frac{\partial}{\partial \nabla}\left(\frac{\partial U}{\partial T}\right)_h. \tag{8}$$

The Helmholtz free energy of the system can be defined as:

$$F = U - TS \tag{9}$$

in which, according to the third law of thermodynamics, it can be written in the form

$$F = U - T\int_0^T \frac{C}{T'} dT'. \tag{10}$$

The second term at the right–hand side of Eq. (10) (the integral which appears) is the entropy of the system according to the second law of the thermodynamics.

## 3. Numerical results and discussions

Some characteristic properties of the mixed spin (1/2-1) HIN system with core-shell structure have examined in this section. Throughout of the paper, $J_C$ have selected as the unit of the system for calculations and it takes 1.0. The effects of the Hamiltonian parameters have studied on the thermodynamic and magnetic properties and hysteresis behaviors of the HIN system.

### 3.1 Thermodynamic and magnetic properties

The magnetizations, susceptibilities, internal energies and free energies of mixed spin (1/2-1) HIN system have presented in Figs. 2, 3 and 4. Figs. 2(a), 3(a) and 4(a) show the thermal variation of the magnetizations for r = 0.5, $\Delta_S$ = -0.5 and D = -1.0, r = -1.0, $\Delta_S$ = -0.5 and D = -1.0, and r = 0.5, $\Delta_S$ = 1.0 and D = -4.0 values, respectively. In Figs. 2(a) and 3(a), the shell and total magnetizations decrease to zero continuously as the temperature increases; therefore, a second-order phase transition occurs at $T_C$ = 1.12 and 1.62, respectively. While the core magnetizations reduce to zero with grow up temperature in Fig. 2(a), it is increase to zero with temperature increase in Fig. 3(a). In Fig. 4(a), the core, shell and total magnetizations undergo a first-order phase transition at $T_t$ = 1.78 along with increase of temperature. The transition is from the ferromagnetic phase to the paramagnetic phase in Fig. 2(a) and 4(a), and from the antiferromagnetic phase to the paramagnetic phase in Fig. 3(a). A few explanatory and interesting results of the core, shell and total susceptibilities are plotted in Figs. 2(b), 3(b) and 4(b) for the above-mentioned values. In Figs. 2(b) and 3(b), at zero temperature, the susceptibilities are equal to zero and with increase of temperature they are increase. As we clearly see from these figures, in the vicinity of the transition temperatures, the susceptibilities increase very rapidly and goes to infinity. Moreover, in the vicinity of $T_t$ the susceptibilities rapidly increases for $T < T_t$ and suddenly decreases for $T > T_t$ in Fig. 4(b). The internal energies are obtained in Figs. 2(c), 3(c) and 4(c) for aforementioned parameters values. From these figures, one can clearly know that the internal energy increases rapidly with the increasing of the temperature. We can see that core, shell and total energy curves have an inflexion point in which the derivative of the internal energy with respect to the temperature is discontinuous. While this point is correspond to a second-order phase transition point for Figs. 2(c) and 3(c), it is illustrate the first-order phase transition point for Fig. 4(c). Figs. 2(d), 3(d) and 4(d) indicate the core, shell and total free energy behaviors for the above mention physical parameters. As known, entropy is not important at low temperatures and ground state energy correspond to the free energy of the system. But, along with the temperature growing, in order to minimize its free energy the system wants to maximize its entropy. In this way, entropy becomes important. Moreover, at the critical temperature in Figs. 2(d) and 3(d), the free energy of the system is continuous. It means that the type of the phase transition is second-order phase transition. On the other hand, For Fig. 4(d), free energy curves stand for an inflexion that it corresponds a discontinuous behavior or a first-order transition.

**3.2 Hysteresis behaviors**

*3.2.1 The influence of the temperature*

In order to investigate the effect of the temperature on the hysteresis behaviors of the HIN system, Fig. 5 is obtained for selected six typical temperature in the case of $r = 0.5$, $\Delta_S = 0.0$ and $D = 0.0$ fixed values. From Fig. 5 we can see that the magnetization curves are symmetric for both positive and negative values of the external magnetic field. We can also see that the hysteresis loops do not occur at temperatures above the transition temperature $T_C = 2.8$ and the type of hysteresis loops becomes narrower as the temperature increases below the transition temperature. Similar behaviors of the hysteresis loops have been observed for nano systems within the EFT [46-50].

*3.2.2 The influence of the single-ion anisotropy, i.e. crystal field*

To better understand the effect of the crystal field on the hysteresis loops, we have plotted the magnetization curves versus the applied field h for $T = 0.5$, $r = 0.5$ and $\Delta_S = 0.0$, and for different values of the crystal field ($D = 0.0, -1.0, -2.0, -3.0$ and $-4.0$) as seen in Fig. 6. At first, we can see that there is only one hysteresis loop in Fig. 6(a) for $D = 0.0$. With decreasing of the crystal field, hysteresis loop area is narrowing. This fact is clearly seen in Fig. 6(b) and 6(c). As from $D = -2.25$ value, the single loop start turn to the double loop and to clearly see double loop we plotted magnetizations curves for $D = -3.0$. As we clearly see from Figs. 6(d) and 6(e), while the crystal field decreases, the hysteresis loops are diverge from each other for a particular range value of the external magnetic field.

*3.2.3 The influence of the ferromagnetic and antiferromagnetic interfacial coupling*

In this part, our investigations will be ferromagnetic and antiferromagnetic. Hence, r> 0 (positive core-shell coupling) and r<0 (negative core-shell coupling) correspond to the ferromagnetic and antiferromagnetic interfacial couplings, respectively. In Fig. 7, we show the dependence of the hysteresis loops of the HIN system at T = 0.2, D = 0.0 for $\Delta_S$ = 0.0, and for various values of interface coupling constant (r = 0.01, 0.25 and 1.0) on the ferromagnetic case. We can see that when the ferromagnetic coupling constant is small, the hysteresis consists two loops as seen in Fig. 7(a). One can see that with the r increases, the hysteresis behavior changes from two loops to one loop and the hysteresis loop area is decreasing. This fact is clearly see from the Figs. 7 (b) and (c). Moreover, the behaviors of the hysteresis loops for antiferromagnetic case is similar to for ferromagnetic case. For instance, the hysteresis consist two loops for r = -0.01 and with the decrease of the r, two loops start turn to the single loop as seen in Figs. 8 (c) and (d).

## 4. Summary and Conclusions

In this paper, we have studied the thermodynamic and magnetic properties (magnetizations, susceptibilities, internal energies and free energies) and hysteresis behaviors of the mixed spin (1/2-1) hexagonal Ising nanowire (HIN) system by using the framework effective-field theory with correlations. It has been shown that the system undergoes a second- and first-order phase transition as depending on the physical parameters of the system. The free energy of the system have investigated to confirm the accuracy of the phase transition points. Our results show that when the temperature increases the hysteresis loop areas decrease and the hysteresis loops disappear at above critical temperature. Moreover, different hysteresis loop behaviors have been observed such as single, double and triple hysteresis loops in the system. Finally, we hope that the study of HIN system may open a new ferrimagnetism as well as new field in the research of magnetism and present work will be potentially helpful for studying higher spins and more complicated nanowire systems.

## References

[1] D. F. Emerich and C. G. Thanos Expert Opin. Biol. Ther. **3,** 655 (2003)
[2] A. Fert, L. Piraux, J. Magn. Magn. Mater. **200,** 338 (1999)
[3] R.H. Kodama, J. Magn. Magn. Mater. **200,** 359 (1999)
[4] J. E. Wegrowe, D. Kelly, Y. Jaccard, Ph. Guittienne, J.Ph. Ansermet, Eur. Phys. Lett. **45,** 626 (1999)
[5] G.V. Kurlyandskaya, M.L. Sanchez, B. Hernando, V.M. Prida, P. Gorria, M. Tejedor, Appl. Phys. Lett. **82,** 3053 (2003)
[6] D.W. Elliott, W.-X. Zhang, Environ. Sci. Technol. **35,** 4922 (2001)
[7] S. Nie, S.R. Emory, Science **275,** 1102 (1997)
[8] R.K. Soong, G.D. Bachand, H.P. Neves, A.G. Olkhovets, H.G. Craighead, C.D. Montemagno, Science **290,** 1555 (2000)
[9] H. Zeng, J. Li, J.P. Liu, Z.L. Wang, S. Sun, Nature **420,** 395 (2002)

[10] V. Skumryev, S. Stoyanov, Y. Zhang, G.C. Hadjipanayis, D. Givord, J. Nogués, Nature, **423**, 850 (2003)

[11] Z. Zhong, D. Wang, Y. Cui, M.W. Bockrath, C.M. Lieber, Science **302**, 1377 (2003)

[12] R. Skomski, J. Phys.: Condens. Matter **15**, 841 (2003)

[13] C. Frandsen, C.W. Ostenfeld, M. Xu, C.S. Jacobsen, L. Keller, K. Lefmann, S. Morup, Phys. Rev. B **70**, 134416 (2004)

[14] Z. Liu, D. Zhang, S. Han, C. Li, B. Lei, W. Lu, J. Fang, C. Zhou, J. Am. Chem. Soc. **127**, 6 (2005)

[15] J. Nogués, J. Sort, V. Langlais, S. Doppiu, B. Dieny, J.S. Munoz, S. Surinach, M.D. Baro, S. Stoyanov, Y. Zhang, Int. J. Nanotechnol. **2,** 23 (2005)

[16] S.I. Denisov, T.V. Lyutyy, P. Hänggi, K.N. Trohidou, Phys. Rev. B **74**, 104406 (2006)

[17] S.I. Denisov, T.V. Lyutyy, P. Hänggi, Phys. Rev. Lett. **97**, 227202 (2006)

[18] X. Zou, G. Xiao, Phys. Rev. B **77**, 054417 (2008)

[19] G. Liu, N. Hoivik, K. Wang, H. Jakobsen, Solar Energy Mater. Solar Cells **105**, 53 (2012)

[20] M. Sebaa, T.Y. Nguyen, R.K. Paul, A. Mulchandani, H. Liu, Mater. Lett. **92**, 122 (2013)

[21] B. Hu, W. Chen, J. Zhou, Sens. Actuators B **176**, 522 (2013)

[22] M. Estrader, A. Lopez-Ortega1, S. Estrade, I.V. Golosovsky, G. Salazar-Alvarez, M. Vasilakaki, K.N. Trohidou, M. Varela, D.C. Stanley, M. Sinko, M.J. Pechan, D.J. Keavney, F. Peiro, S. Surinach, M.D. Baro and J. Nogués, Nature Commun. **4,** 2960 (2013)

[23] T. Kaneyoshi, E.F. Sarmento, I.F. Fittipaldi, Phys. Stat. Sol. B **150,** 261 (1988)

[24] T. Kaneyoshi, J.C. Chen, J. Magn. Magn. Mater. **98,** 201 (1991)

[25] J.A. Plascak, Physica A **198,** 665 (1993)

[26] G.M. Zhang, C.Z. Yang, Phys. Rev. B **48,** 9452 (1993)

[27] G.M. Buendia, M.A. Novotny, J. Zhang, in: D.P. Landau, K.K. Mon, H.B. Schuttler (Eds.), Springer Proceeds in Physics 78, Computer Simulations in Condensed Matter Physics, Vol. VII, Springer, Heidelberg, 223 (1994)

[28] A. Bobak, M. JaVsVcur, Phys. Rev. B **51,** 533 (1995)

[29] Z.H. Xin, G.Z. Wie, T.S. Liu, J. Magn. Magn. Mater. **176,** 206 (1997)

[30] Z.H. Xin, G.Z. Wei, T.S. Liu, Physica A **248,** 442 (1998)

[31] S.L. Yan, C.Z. Yang, Phys. Rev. B **57,** 3512 (1998)

[32] N. Benayad, A. Dakhama, A. Fathi, R. Zerhouni, J. Phys.: Condens. Matter. **10**, 314 (1998)

[33] A. Zaim, M. Kerouad, Y. E. Amraoui, J. Magn. Magn. Mater. **321,** 1077 (2009)

[34] N. Şarlı, Physica B **411,** 12 (2013)

[35] E. Kantar, Y. Kocakaplan, Solid State Commun. **177**, 1 (2014)

[36] H.W. Wu, C.J. Tsai, L.J. Chen, Appl. Phys. Lett. **90,** 043121 (2007)

[37] T.G. Sorop, K. Nielsch, P. Göring, M. Kröll, W. Blau, R.B. Wehrspohn, U. Gösele, L.J. de Jongh, J. Magn. Magn. Mater. **272,** 1656 (2004)

[38] J. Curiale, R.D. Sanchez, H.E. Troiani, A.G. Leyva, P. Levy, Appl. Phys. Lett. **87,** 043113 (2005)

[39] G. Kamalakar, D.W. Hwang, L.P. Hwang, J. Mater. Chem. **12,** 1819 (2002)

[40] D.L. Peng, X. Zhao, S. Inoue, Y. Ando, K. Sumiyama, J. Magn. Magn. Mater. **292,** 143 (2005)

[41] C.-S. Chen, T.-G. Liu, X.-H. Chen, L.-W. Lin, Q.-C. Liu, Q-Xia, Z.-W. Ning, Trans. Nonferr. Met. Soc. China **19,** 1567 (2009)

[42] J.A. Garcia, E. Bertran, L. Elbaile, J. Garcia-Cespedes, A. Svalov, Phys. Status Solidi c **7,** 2679 (2010)

[43] B.Z. Mi, H.Y. Wang, Y.S. Zhou, J. Magn. Magn. Mater. **322,** 952 (2010)

[44] Z. Zhu, Y. Sun, Q. Zhang, J.-M. Liu Phys. Rev. B, **76,** 014439 (2007)

[45] O. Yalçın, R. Erdem, S. Övünç Acta Phys. Pol. A, **114,** 835 (2008)

[46] M. Keskin, N. Şarlı, B. Deviren, Solid State Commun. **151**, 1025 (2011)

[47] A. Zaim, M. Kerouad, M. Boughrara, J. Magn. Magn. Mater. **331**, 37 (2013)

[48] Y. Kocakaplan, E. Kantar, M. Keskin, Eur. Phys. J. B, **86,** 420 (2013)

[49] E. Kantar, B. Deviren, M. Keskin, Eur. Phys. J. B **86,** 253 (2013)

[50] Y. Kocakaplan, E. Kantar Chin. Phys. B **23,** 046801 (2014)

**List of the figure captions**

**Fig. 1.** (Color Online) The schematic representation of a hexagonal Ising nanowire. The blue and red spheres indicate magnetic atoms at the surface shell and core, respectively.

**Fig. 2.** (Color Online) The effects of the physical parameters of the mixed spin (1/2-1) HIN system on: **(a)** magnetizations, **(b)** susceptibilities, **(c)** internal energies and **(d)** free energies for $r = 0.5$, $= \Delta_S = -0.5$ and $D = -1.0$ values.

**Fig. 3.** (Color Online) Same as with Fig. 2, but for $r = -1.0$, $= \Delta_S = -0.5$ and $D = -1.0$ values.

**Fig. 4.** (Color Online) Same as with Fig. 2, but for $r = 0.5$, $= \Delta_S = 1.0$ and $D = -4.0$ values.

**Fig. 5.** (Color Online) The effects of the temperature on the hysteresis behaviors for $r = 0.5$, $\Delta_S = 0.0$ and $D = 0.0$ fixed values and $T = 0.5, 1.0, 1.5, 2.0, 2.5$ and $3.0$.

**Fig. 6.** (Color Online) The effects of the crystal field on the hysteresis loops for T = 0.5, r = 0.5 and Δs = 0.0 fixed values, and for (a) D = 0.0, (b) D = -1.0, (c) D = -2.0, (d) D = -3.0 and (e) D = -4.0.

**Fig. 7.** (Color Online) The effects of the ferromagnetic interface coupling on the hysteresis loops for T = 0.2, D = 0.0 and Δs = 0.0 fixed values, and for (a) r = 0.01, (b) r = 0.25 and (c) r = 1.0.

**Fig. 8.** (Color Online) The effects of the antiferromagnetic interface coupling on the hysteresis loops for T = 0.2, D = 0.0 and Δs = 0.0 fixed values, and for (a) r = -0.01, (b) r = -0.25, (c) r = -0.5 and (d) -0.75.

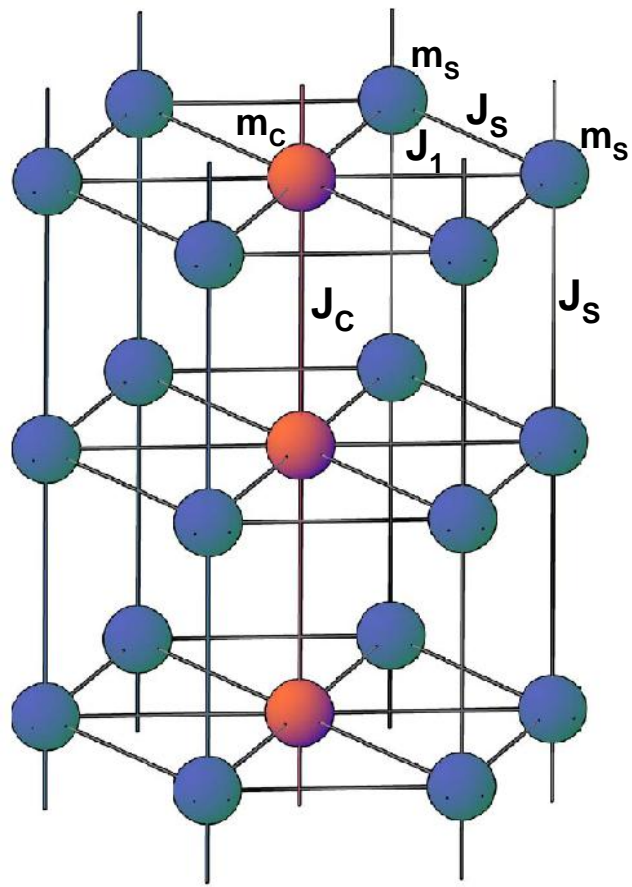

Fig. 1

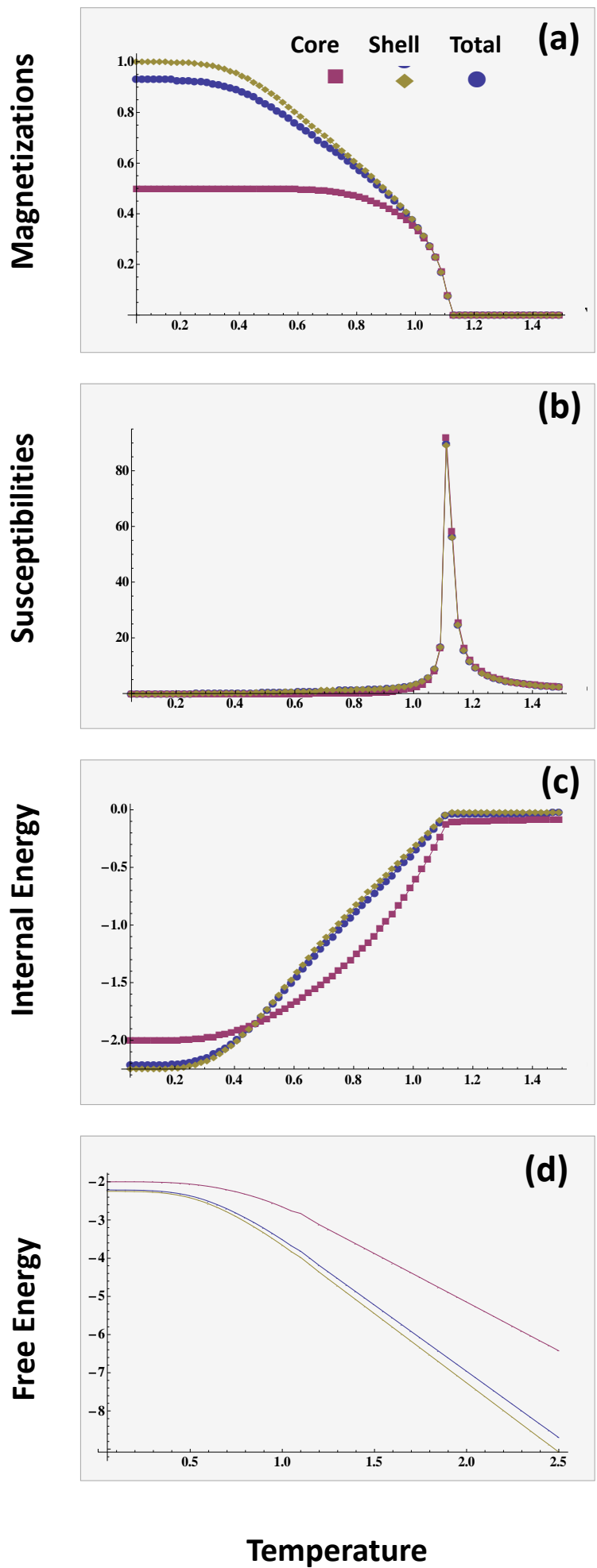

Fig. 2

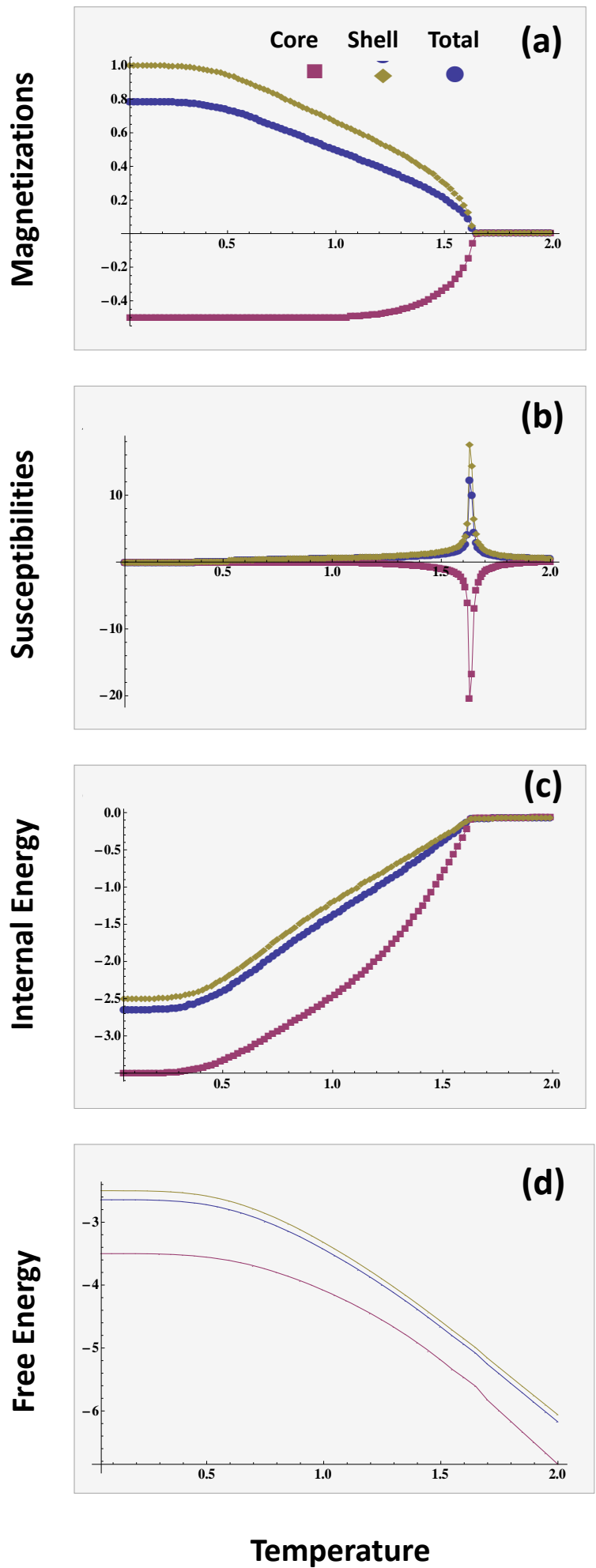

Fig. 3

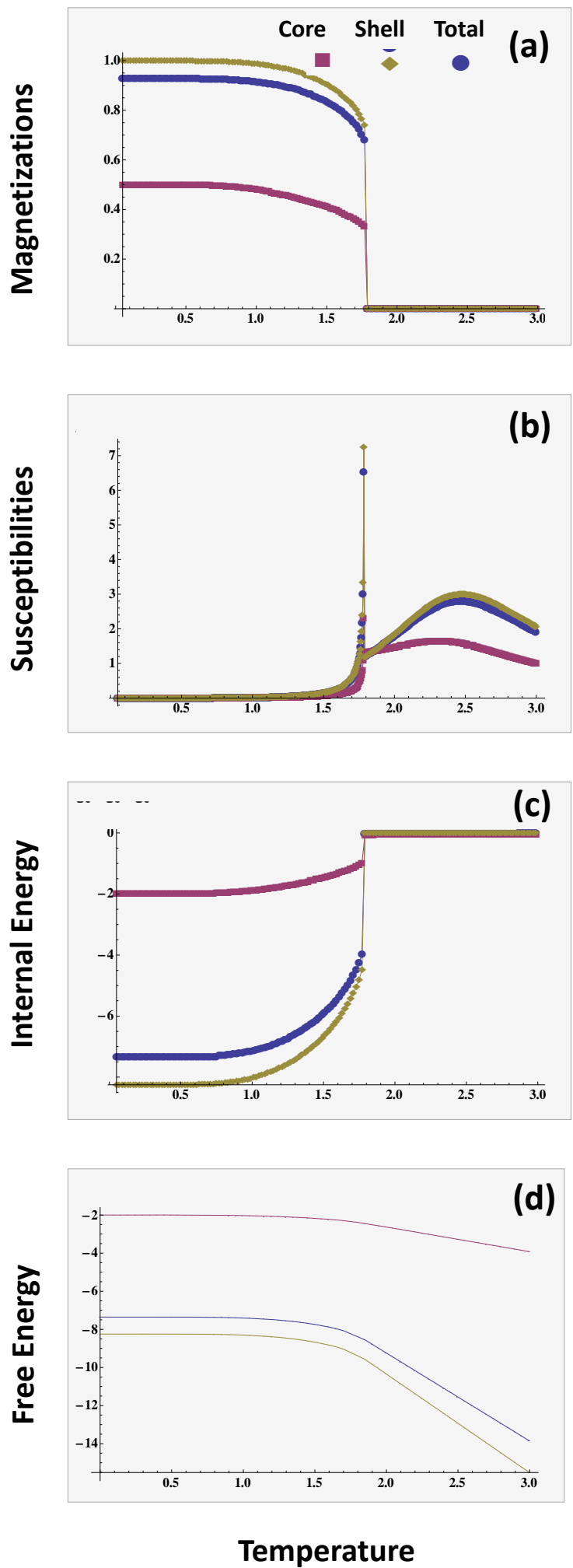

Fig. 4

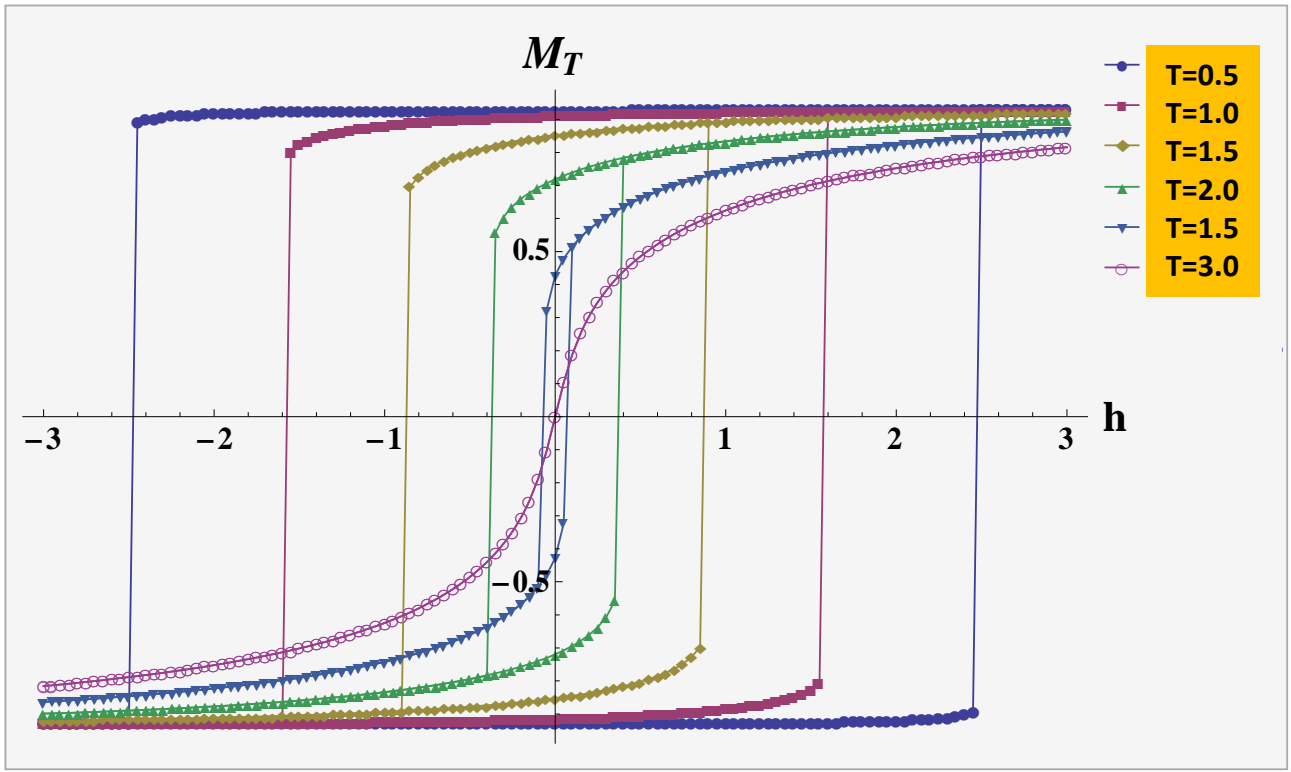

Fig. 5

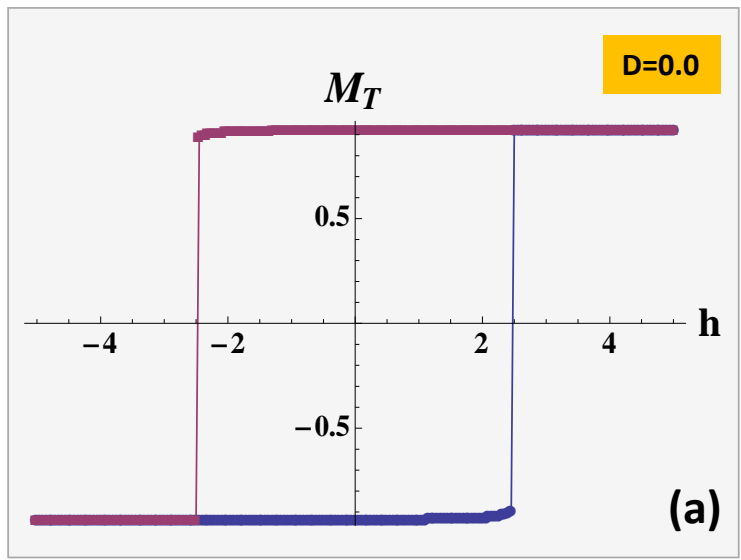
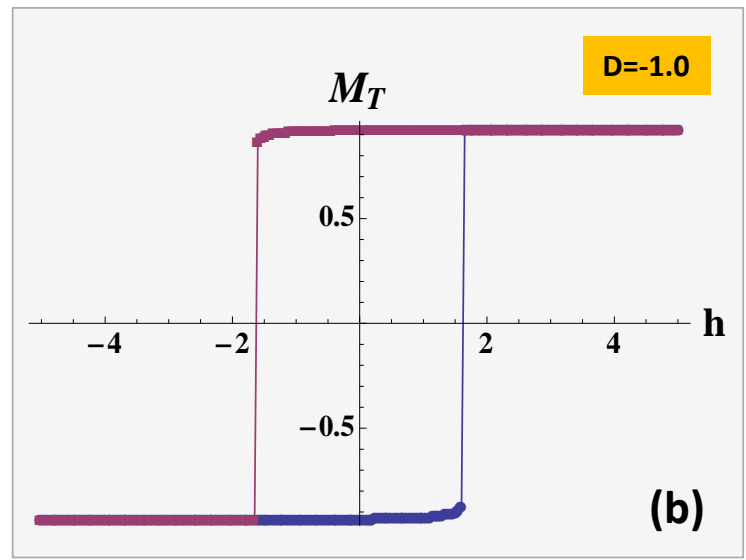
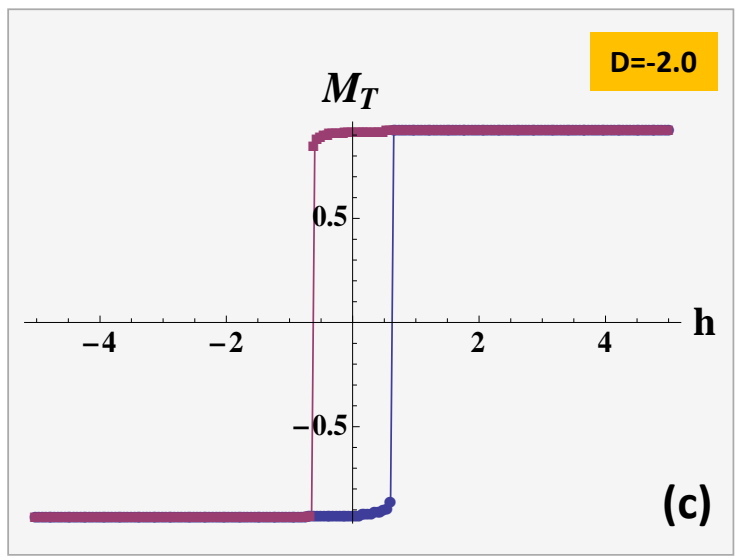
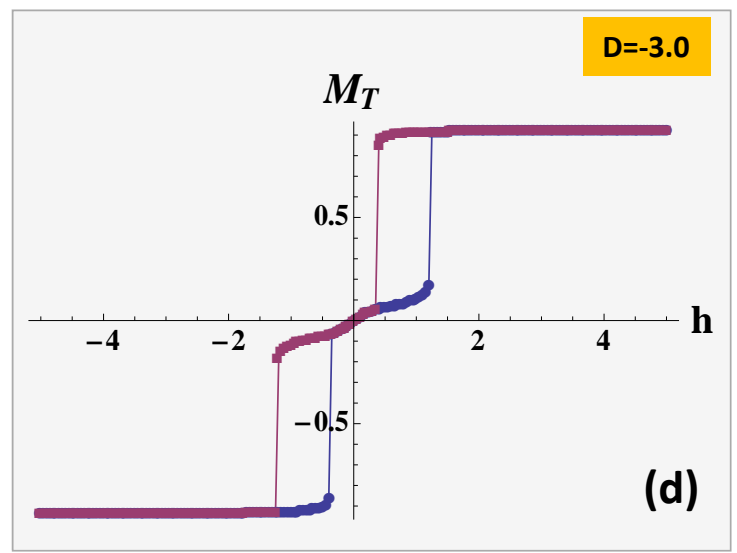
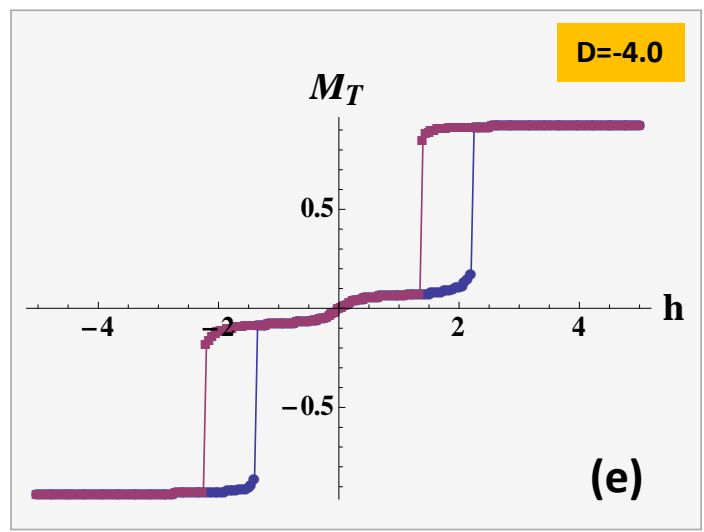

Fig. 6

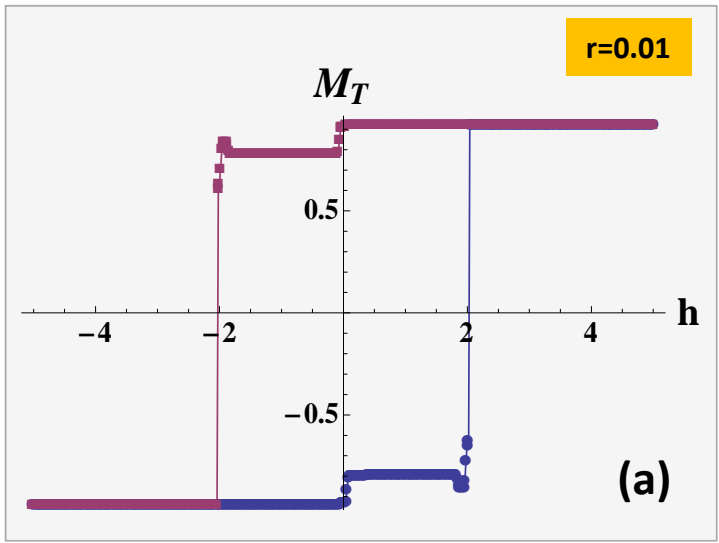 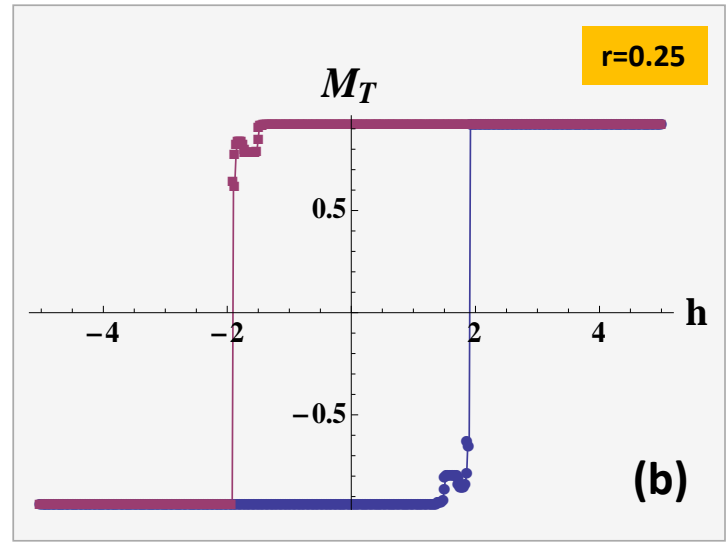

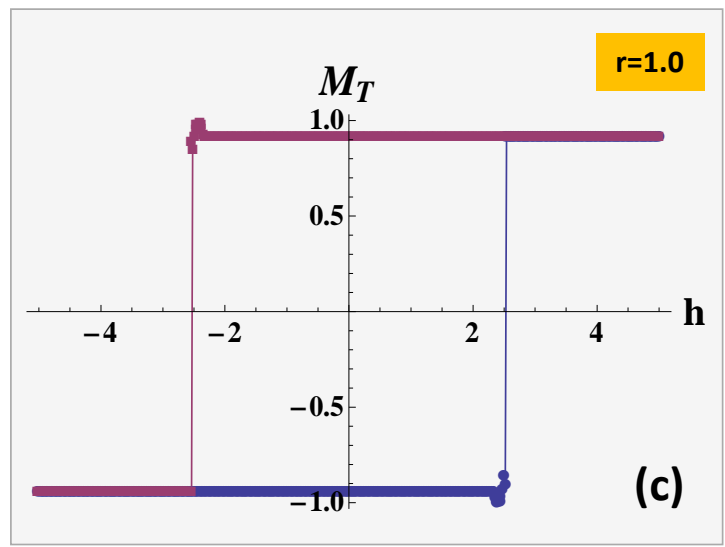

Fig. 7

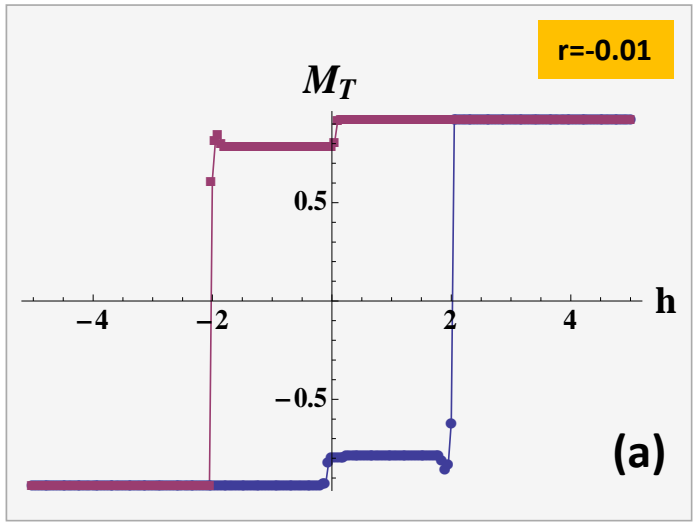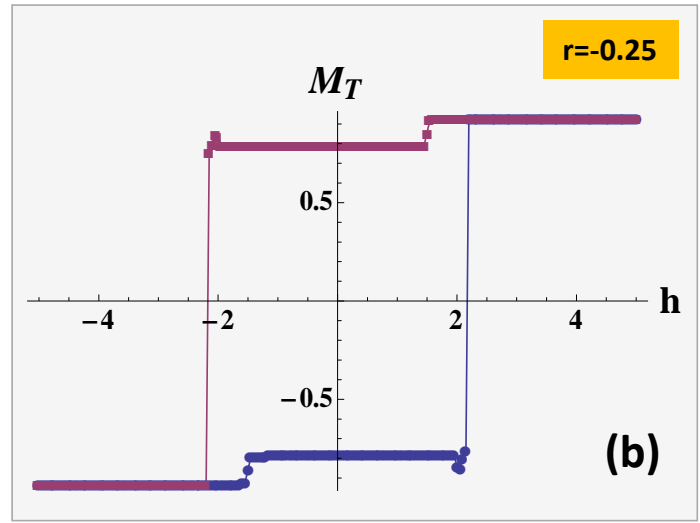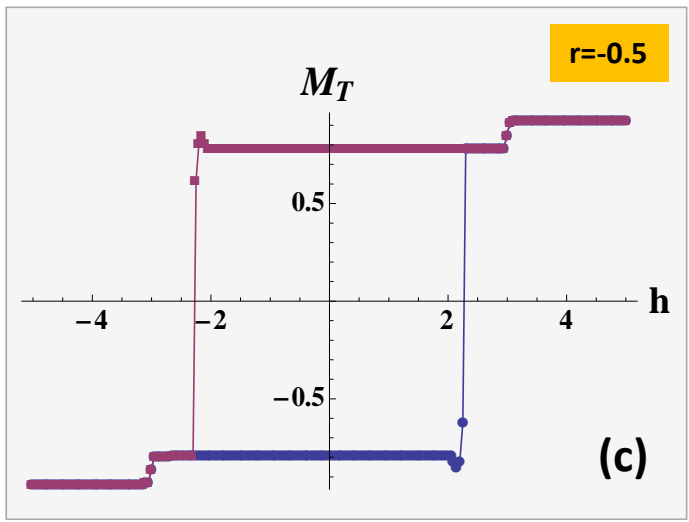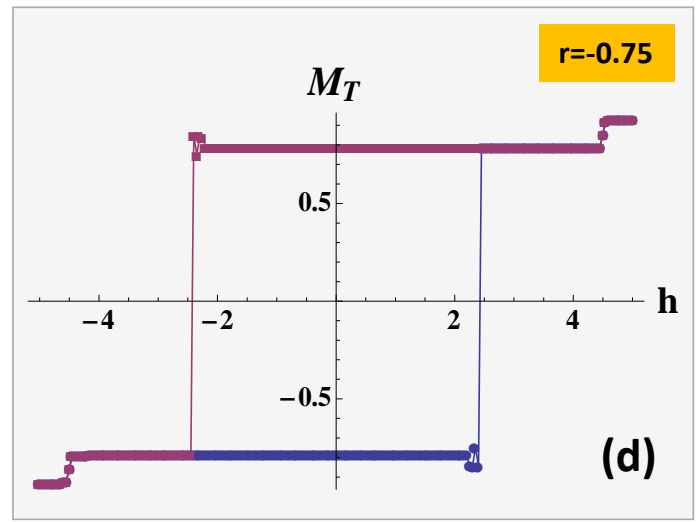

Fig. 8